\documentclass[12pt]{article}
\hoffset = - 0.5 truecm \voffset=-1.4 truecm
\def\beq{\begin{equation}}   \def\eeq{\end{equation}}
\def\bea{\begin{eqnarray}}  \def\eea{\end{eqnarray}} \def\nn{\nonumber}
\def\noi{\noindent} \def\beeq{\begin{eqnarray}}
\def\eeeq{\end{eqnarray}}
\def\lsim{\raise0.3ex\hbox{$<$\kern-0.75em\raise-1.1ex\hbox{$\sim$}}}
\def\gsim{\raise0.3ex\hbox{$>$\kern-0.75em\raise-1.1ex\hbox{$\sim$}}}

\usepackage{graphicx}
\usepackage{epsfig}
\begin{document}

\begin{titlepage}
 
\begin{flushright}
LAPTH 1020/03 \\
IMSc-2003/11/35 \\
LPT-Orsay 03-97\\
IISc-CTS/18/03 \\
December 2003
\end{flushright}
\vspace{1.cm}

\begin{center}
\vbox to 1 truecm {}
{\large \bf An NLO calculation of the electroproduction\par \vskip 3 truemm
   of large-E$_{\bot}$ hadrons}
\par \vskip 3 truemm
\vskip 1 truecm {\bf P. Aurenche$^{a)}$, Rahul Basu$^{(b)}$, M.
Fontannaz$^{(c)}$, R.M. Godbole$^{(d)}$} \vskip 3 truemm

{\it $^{(a)}$ LAPTH, UMR 5108 du CNRS associ\'ee \`a l'Universit\'e
de Savoie, \\ BP 110, Chemin de Bellevue, 74941 Annecy-le-Vieux
Cedex, France}

\vskip 3 truemm

{\it $^{(b)}$ The Institute of Mathematical Sciences,\\ Chennai 600 113, India}

\vskip 3 truemm

{\it $^{(c)}$ Laboratoire de Physique Th\'eorique, UMR 8627 CNRS,\\
Universit\'e Paris XI, B\^atiment 210, 91405 Orsay Cedex, France}\\

\vskip 3 truemm

{\it $^{(d)}$ Centre for Theoretical Studies,\\ Indian Institute of
Science, Bangalore 560012, India}\\

\vskip 2 truecm

\normalsize

\begin{abstract}
We present a Next-to-Leading Order calculation of the cross section for 
the leptoproduction of large-$E_{\bot}$ hadrons and we compare our
predictions with H1 data on the forward production of $\pi^0$. We find
large higher order corrections and an important sensitivity to the
renormalization and factorization scales. These large corrections are
shown to arise in part from BFKL-like diagrams at the lowest
order.\end{abstract}

\end{center}

\end{titlepage}
\baselineskip = 22 pt
\section{Introduction}
\hspace*{\parindent}
The electroproduction of large-$E_{\bot}$ hadrons which is observed by
the HERA experiments H1 \cite{1r,2r,2bisr} and ZEUS \cite{3r} may provide
important tests of QCD. In addition to the study of the partonic
subprocesses, of the parton distributions in the proton and of the
fragmentation functions, it also offers the possibility to observe the
virtual photon structure function. A contribution of the latter is
indeed expected when the hadron transverse energy squared $E_{\bot}^2$ is
much larger than the photon virtuality $Q^2 = |q^2|$~; in this case the
virtual photon structure function contribution, proportional to
$\log{E_{\bot}^2 \over Q^2}$, can be important. \par

Another interest of this reaction is the study of the production
mechanisms of forward hadrons. Indeed, the forward region can be
associated with BFKL dynamics and provide tests of the latter
\cite{4r}. Several papers have studied the production of
large-$E_{\bot}$ jets \cite{5r} and hadrons \cite{6r} in the forward
direction, and have concluded that these reactions are relevant for the
study of BFKL dynamics. However, results from H1 \cite{2r}
show that theoretical predictions based on the DGLAP dynamics,
implemented in RAPGAP \cite{7r} or based on BFKL dynamics
\cite{8r} are both in agreement with data. These theoretical results do
not necessarily contradict each other, since the same Feynman graphs may
contribute to both of them. But it is clear that a deeper understanding
of the underlying forward dynamics requires quantitative predictions,
which are not fully realized in the existing literature above. Ref. \cite{7r}
which implements the DGLAP dynamics rests on a Leading Order (LO) 
approximation and suffers
from scale dependence which forbids an absolute normalization. Ref.
\cite{8r} takes into account some Higher Order (HO) corrections to the LO BFKL
equation, but does not include non-BFKL contributions. \par

In this paper we calculate the HO corrections to the Born subprocesses
associated with the electroproduction of large-$E_{\bot}$ hadrons
namely the QCD Compton process $\gamma^*+q \to g+q$ and the fusion
process $\gamma^* + g \to q + \bar{q}$. The Born subprocess cross
sections, of order ${\cal O}(\alpha\alpha_s)$, and the HO subprocess
cross sections, of order ${\cal O}(\alpha\alpha_s^2)$, are convoluted
with parton distributions and fragmentation functions calculated at the
NLO approximation. The total NLO cross-section (Born + HO
contributions) is sensitive to the choice of the renormalization and
factorization scales, but there is a compensation between the
variations of the Born and HO contributions in such a way that the NLO
cross section is more stable than the LO (containing only the Born
terms) cross sections. \par

The work presented here is a fixed order (for the subprocess
cross section), NLO calculation, the DGLAP dynamics being included by
the scale dependent distributions and fragmentation functions. However,
among the HO contributions, one of them corresponds to the lowest order
BFKL cross section, namely, the reaction $\gamma^* + g \to g + q +
\bar{q}$. As this contribution is part of an HO calculation, we have a
way to establish a link between the normalization of our NLO cross
section and that of the (lowest order) BFKL cross section. \par

As mentioned in the beginning of this section, the HO calculation
generates a contribution proportional to the virtual photon structure
function. At order ${\cal O}(\alpha\alpha_s^2)$ we obtain the
Born expression of this structure function, proportional to
$\log E_{\bot}^2/Q^2$ and we shall study how important this contribution to
the large-$E_{\bot}$ forward hadron cross section is in the H1
kinematical configuration. All order contributions to the virtual
photon structure function can be resummed using an inhomogeneous DGLAP
evolution equation. Here we shall briefly discuss this possibility,
leaving for another publication \cite{9r} a detailed analysis of the
resummed virtual photon structure function. \par

As a final point concerning the nature of this calculation, let us
emphasize the fact that it describes the production of large-$E_{\bot}$
hadrons in the virtual-photon proton centre-of-mass system (CMS) and that it
is also valid in the limit $Q^2 = 0$, the large scale then being provided by
$E_{\bot}^2$. It must be compared with experimental results which impose
a lower bound on the final hadron $E_{\bot}$ in the $\gamma^*-P$ CMS,
as is done by the H1 and ZEUS collaboration \cite{2r,2bisr,3r}. Therefore
the present work does not consider the target fragmentation mechanism,
which requires the introduction of fracture functions
\cite{10r,11r,12r,13r}. It is also different from the inclusive calculations of
ref. \cite{12r,13r} in that it is an exclusive NLO partonic generator. This
allows us to calculate various types of correlations (for example
between large-$E_{\bot}$ hadron and jets) and facilitates the
implementation of experimental cuts. \par

In the next section we shall present an overview of the relevant DIS
kinematics and of the method used to calculate the HO corrections.
Section 3 is devoted to a discussion of the virtual photon structure
function and, in section 4, we compare our theoretical results with H1
data \cite{2r}. We shall discuss in detail the importance of the
virtual photon contribution and of the BFKL-like contribution in the H1
kinematical region. Section 5 studies the production of
large-$E_{\bot}$ hadrons in the central region in rapidity. Section 6
is the conclusion.

\section{The NLO calculation}
\hspace*{\parindent}
The kinematics of the reaction $e(\ell ) + p (P)\to e (\ell ') +
h(P_4) + X$ is fixed
by the observation, in the laboratory frame, of the outgoing lepton and
hadron $h$ momenta \cite{2r}. We define the photon variables (axis
$Oz$ along the initial proton momentum)

\bea \label{1e} &&Q^2 = - q^2 = - (\ell - \ell ')^2 \nn \\ &&y \equiv
{q^{(-)} \over \ell^{(-)}} = {q^0 - q^z \over \ell^0 - \ell^z} =
{P \cdot q \over P\cdot \ell} = {Q^2 \over S} \ {1 \over x_{Bj}} ,\eea

\noi where we have neglected the proton mass and used the notation $S =
(\ell + P)^2$ and $x_{Bj} = {Q^2 \over 2P\cdot q}$. The outgoing hadron
is defined by its transverse energy $E_{\bot 4}^{Lab}$ and its pseudo
rapidity $\eta_4^{Lab}$.

The inclusive cross section is written in terms of the leptonic
tensor (summed and averaged over the spins of the leptons)

\beq \label{2e} \ell^{\mu\nu} = 2 \left ( \ell^{\mu} \ell '^{\nu} +
\ell'^{\mu} \ell^{\nu} - g^{\mu\nu} \left ( \ell \cdot \ell ' - m_e^2
\right ) \right ), \eeq

\noi and the hadronic tensor $T_{\mu\nu}$ which describes the
photon-proton collision

\beq \label{3e} {d\sigma \over d\varphi dQ^2 dy} = {\alpha \over 2 \pi}\
{1 \over 2 \pi} \ {1 \over 2S} \int {1 \over 2} \ {\ell^{\mu\nu} T_{\mu\nu}
\over q^4} dPS, \eeq

\noi where

$$dPS = (2 \pi )^4 \delta^4 \left ( q + P - \sum_{i=1}^n  p_i\right )
\prod_{i=1}^n {d^4p_i
\over (2 \pi )^3} \delta (p_i^2) \theta (p_i^0)$$

\noi is the final state hadron phase space element and $\varphi$ the photon
azimuthal angle. (A sum over the number of final hadrons is understood
in (\ref{3e})).\par

The hadronic tensor can be calculated as a convolution between the
partonic tensor $t_{\mu\nu}$ which describes the interaction between the
virtual photon and the parton of the proton, and the parton
distribution in the proton $G_a(x,M)$. The fragmentation of the final
parton which produces a large-$E_{\bot}$ hadron is described by the
fragmentation function $D_b^h (z, M_F)$. These distributions depend on
the factorization scales $M$ and $M_F$,

\beq \label{4e} \int T_{\mu\nu} dPS = \sum_{a,b} \int {dx \over x}
G_a(x,M) \int
dz \ D_b^h(z, M_F) t_{\mu\nu}^{ab} \cdot dps ,\eeq

\noi where $dps$ is the phase space element of the partons produced in
the hard photon-parton collision. From expressions (\ref{3e}) and
(\ref{4e}), we obtain

\bea \label{5e} &&{d\sigma \over d\varphi dQ^2 dydE_{\bot 4}^{Lab}
d\eta_4^{Lab}} = {E_{\bot 4}^{Lab} \over 2 \pi} {\alpha \over 2 \pi} \sum_{a,b}
\int dx G_a(x,M) \int {dz \over z^2} D_b^h(z,M_F) \nn \\
&&\int
{d\varphi_4^{Lab} \over 2 \pi } {1 \over (4 \pi)^2} {1 \over 2 xS}
{\ell^{\mu\nu} t_{\mu\nu}^{ab} \over q^4} dps', \eea

\noi where the phase space $dps'$ no longer contains parton 4
which fragments into $h(P_4)$. Up to this point we have been writing
$E_{\bot 4}^{Lab}$ and $\eta_4^{Lab}$ to emphasize the frame in which
the reaction is observed. Of course (\ref{5e}) is valid in any frame
and from now on we shall drop the index ``Lab''. \par

The tensor product is a series in
$\alpha_s$. Taking into account the first and second order
contributions, we rewrite (\ref{5e}) as

\bea \label{6e} &&{d\sigma \over d\varphi dQ^2 dy dE_{\bot 4} d\eta_4}
= {\alpha \over 2 \pi} \sum_{a,b} \int dx G_a(x, M) \int {dz \over z^2}
D_b^h(z,M_F) \nn \\ &&\left \{ {\alpha_s(\mu ) \over 2 \pi} \
{d\widehat{\sigma}_{a,b}^{Born}(x, z)  \over d \varphi dQ^2 dy dE_{\bot
4} d\eta_4} + \left ( {\alpha_s (\mu) \over 2 \pi}\right )^2\
{dK_{ab}^{HO}(x,z,\mu,M,M_F) \over d \varphi dQ^2 dy dE_{\bot 4}
d\eta_4} \right \} \ . \eea

\noi The cross sections $\widehat{\sigma}_{ab}^{Born}$ are the
subprocess Born cross sections which describe the electroproduction of
a large-$E_{\bot}$ parton $b$ and $K_{ab}^{HO}$ are the associated
Higher Order corrections. \par

Figure 1a shows the Born term corresponding to the QCD Compton (QCDC) process
(here partons $a$ and $b$ are quarks). When $a$ is a gluon, we have the
so called photon-gluon fusion Born term. Examples of a graph
contributing to HO corrections to the QCDC term are shown in Fig. 1b,
c. (The numbers 1 to 5 label the initial and final partons according to
a convention used in the HO calculations described below). \par

\begin{figure}
\centering
\includegraphics[width=5in, height=2in]{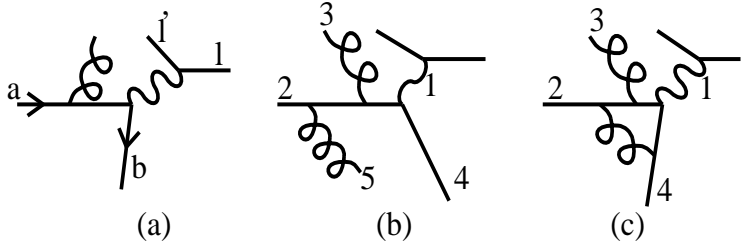}
\caption{The QCD Compton subprocess (Fig. 1a) and real (virtual) HO
corrections to it (Fig. 1b (Fig. 1c)).}
\end{figure}

In expression (\ref{6e}) we have explicitly written the dependence of
the cross section on the strong coupling constant $\alpha_s(\mu)$ which
depends on the renormalization scale~$\mu$.\par

It is convenient to perform the calculation of
$\widehat{\sigma}^{Born}$ and $K^{HO}$ in the virtual photon-proton
center of mass system, and from now on we shall work in this frame.
In fact, the H1 collaboration explicitly uses this frame to place cuts
on the outgoing hadron transverse momentum.
We take the positive $Oz$ axis along the proton momentum (as per the H1
convention) and the leptons are contained in the $Oxz$-plane. \par

It is instructive to give a more explicit form to the tensor product
in the $\gamma^*-p$ frame
using expression (\ref{2e}) and defining the transverse polarization
vectors $\varepsilon_1^{\mu} = (0,1,0,0)$, $\varepsilon_2^{\mu} =
(0,0,1,0)$ and the scalar polarization vector $\varepsilon_s^{\mu} = {1
\over \sqrt{Q^2}} (q^z, 0, 0, q^0)$ with $q^{\mu} = (q^0, 0, 0, q^z)$ being
the virtual photon momentum

\bea
\label{7e}
&&\ell^{\mu\nu} t_{\mu\nu} = Q^2(t_{11} + t_{22}) + 4 \left ( {Q^2(1
- y) \over y^2} - m_e^2 \right ) t_{11}\nn \\
&&+ 4 {2 - y \over y} \ell_x \sqrt{Q^2} \ t_{s1} + Q^2 {4 (1 - y)
\over y^2} t_{ss},
\eea

\noi where $y \equiv {q^0 - q^z \over \ell^0 - \ell^z} = {P \cdot q \over P
\cdot \ell}$ is identical to the variable defined in (\ref{1e}) in
the Lab frame. \par

In the limit $Q^2 \to 0$ and after azimuthal averaging over $\varphi_4$
we recover the
unintegrated Weizs\"acker-Williams expression

\beq \label{8e} {1 \over 2} \ {\ell^{\mu\nu} t_{\mu\nu} \over Q^4} =
\left ( {1 + (1 - y)^2 \over y Q^2} - {2y \ m_e^2 \over Q^4}\right )
\sigma_{\bot} +
{\cal O}\left ( \left ( Q^2\right )^0 \right ) \eeq

\noi with $\sigma_{\bot} = {1 \over 2y} \left ( t_{11} + t_{22}\right
)$. \par

Actually the limit (\ref{8e}) is correct only if $\lim\limits_{Q^2
\to 0} t_{ss} =
{\cal O}(Q^2)$. This is not true if an initial collinearity is present in
the partonic tensor (light partons are massless) which leads to the
behavior $\lim\limits_{Q^2 \to 0} t_{ss} = {\cal O}(1)$. This point will be
discussed in a forthcoming publication \cite{9r}.\par

After these kinematical preliminaries, let us describe the calculation
of the HO corrections which uses the phase-space slicing method elaborated in
ref. \cite{14r}. We outline the strategy only briefly; for more details
we refer to \cite{15r}.\par

For a generic reaction $1 + 2 \to 3 + 4 + 5$ (fig. 1), at least two
particles of the final state, say 3 and 4, have a high $E_{\bot}$ and
are well separated in phase space, while the last one, say 5, can be
soft, and/or collinear to any of the four others. Of course on the
photon side there is no collinear divergence as long as $Q^2$ is
different from zero. This part of the calculation is related to the
virtual photon structure function to be discussed below. In order to
extract the other singularities, the phase space is cut into two
regions~: \par

(1) Part I where the norm $E_{\bot 5}$ of the transverse momentum of
particle 5 is required to be less than some arbitrary value $E_{\bot
m}$ taken to be small compared to the other transverse momenta. This
cylinder contains the infrared and the initial state collinear
singularities. It also contains a small fraction of the final state
collinear singularities.\par

(2) Parts IIa(b) where the transverse momentum vector of particle 5 is
required to have a norm larger than $E_{\bot m}$, and to belong to a
cone $C_3(C_4)$ about the direction of particle 3(4), defined by
$(\eta_5 - \eta_i)^2 + (\phi_5 - \phi_i)^2 \leq R_{th}^2$ ($i = 3, 4)$,
with $R_{th}$ some small arbitrary number. $C_3(C_4)$ contains the
final state collinear singularities appearing when 5 is collinear to
3(4).\par

(3) Part IIc where $E_{\bot 5}$ is required to have a norm larger than
$E_{\bot m}$, and to belong to neither of the two cones $C_3$, $C_4$.
This slice yields no divergence, and can thus be treated directly in 4
dimensions. For this regular part of the calculations we use the cross
sections from  \cite{16r}.\par

The contributions from regions I and IIa, b are calculated analytically
in $d = 4 - 2 \varepsilon$ dimensions and then combined with the
corresponding virtual corrections (borrowed from DISENT \cite{16r}) such
that the infrared singularities cancel each other, leaving only the initial
(final) state collinear singularities, which are
factorized and absorbed into the parton distribution (fragmentation)
functions. The $\overline{MS}$ factorization and
renormalization schemes are used in this calculation.\par

After the cancellation, the finite remainders of the soft and collinear
contributions in parts I and IIa, b, c separately depend on large
logarithms $\ln \ E_{\bot m}$, $\ln^2 \ E_{\bot m}$ and $\ln \
R_{th}$. When combining the different parts, the
cancellations of the $E_{\bot m}$ and $R_{th}$ dependent terms occur. 
Actually, in part I, the finite terms are approximated by collecting 
all the
terms depending logarithmically on $E_{\bot m}$ and neglecting the
terms proportional to powers of $E_{\bot m}$. Similarly in parts IIa
and IIb we keep only the logarithmic terms $\ln \ R_{th}$. Therefore the
parameter $E_{\bot m}$ must be chosen small enough with respect to
$E_{\bot 4}$ so that the neglected terms can be safely 
dropped. On the other hand, it cannot be chosen too small for then
numerical instabilities may occur. Similar remarks are also valid
for the $R_{th}$ cut-off.\par

This approach allows us to build a partonic event generation which is
very flexible~; various sorts of observables and experimental cuts
being easily handled. More references to this method, which has been
used to calculate NLO correction to several photoproduction and
hadroproduction reactions can be found in ref. \cite{17r}.

\section{The resolved contribution}
\hspace*{\parindent}
As mentioned in the Introduction, the calculation of the HO 
corrections leads to a contribution
proportional to $\log{E_{\bot 4}^2 \over Q^2}$ (when $E_{\bot 4}^2 
\gg Q^2$), the so-called
resolved photon contribution. Indeed, let us consider the contribution
associated with Fig. 1b in which we interchange the label 5 and 4. The
integration over $E_{\bot 5}$ (the unobserved final quark momentum)
produces, among other contributions, a logarithm contribution
$\log{E_{\bot 4}^2 \over Q^2}$ associated with a configuration in
which the final quark is collinear to the virtual photon. More
explicitly, we obtain the following expression (for a transversely
polarized photon)

\bea \label{10e} \sigma_{\bot} &=& {\alpha \over 2 \pi} \int dz \left [
z^2 + (1-z)^2 \right ] \left \{ \int_{0}^{E_{\bot 4}^2}
{dk_{\bot}^2 \over k_{\bot}^2-q^2z(1-z)} \right .\nn \\
&&\left . + z(1-z) q^2 \int_{0}^{E_{\bot 4}^2} {dk_{\bot}^2 \over 
(k_{\bot}^2-q^2z(1-z))^2}
\right \} \widehat{\sigma}(0) dps ,\eea

\noi in which we have defined $k = q - p_5$ and $z = {k^{(-)} \over
q^{(-)}}$. The cross section of the $2 \to 2$ subprocess is calculated
with $k$ on-shell ($k^2 = 0$) and $dps$ is the final partonic phase
space of parton 3 and 4. This expression, as well as a similar one
for the scalar cross section, will be derived in ref.\cite{9r}. A 
discussion
of the second term of (\ref{10e}) (the non-logarithmic piece) is also 
postponed
to this paper. Here we are interested in the term proportional to $\log
E_{\bot 4}^2/Q^2$ and in a discussion of the upper limit of the
integral in (\ref{10e}).


 From (\ref{10e}) we obtain the term

\beq \label{11e} \sigma_{\bot} \simeq \int dz P_{q\gamma}(z) \left \{ \log \
{E_{\bot 4}^2 \over Q^2}  \right \}
\widehat{\sigma}(0)\ dps ,\eeq

\noi (with the definition $P_{q\gamma}(z) = {\alpha \over 2 \pi} (z^2 +
(1-z)^2)$ which defines the quark distribution in the
virtual photon

\beq \label{12e} q_{\gamma} (z, E_{\bot 4},Q^2) = P_{q\gamma}(z) \ \log
\ {E_{\bot 4}^2 \over Q^2} \ .\eeq

\noi In
expression (\ref{12e}), the quark distribution is calculated with no
QCD correction. However when $E_{\bot 4}^2 \gg Q^2$ it becomes
important to calculate these corrections and to replace (\ref{12e}) by
the LO or NLO expressions of the quark distribution \cite{9r,A,B,C}.

In this paper we are only interested in the study of the importance of
the resolved contribution obtained in the kinematical configuration of
the H1 experiment and we content ourselves with the lowest order
expression (\ref{12e}). However this expression, obtained in the limit
$E_{\bot 4}^2 \gg Q^2$, must be elaborated in order to cover other
kinematical configurations as well. When $Q^2 \ \gsim\ E_{\bot 4}^2$,
we cannot neglect the $k_{\bot}^2$ dependence of
$\widehat{\sigma}(k_{\bot}^2)dps$ which suppresses the logarithmic
integration in expression (\ref{10e}). The resolved
contribution coming from the collinear configuration can be
approximated by the form

\beq \label{neweq} \sigma_{\bot} \simeq \int dz \ P_{q\gamma}(z) \log
{Q^2 + E_{\bot 4}^2 \over Q^2} \ \widehat{\sigma}(0)\ dps, \eeq

\noi which has the correct limit for $E_{\bot 4}^2 \gg Q^2$ and
$E_{\bot 4}^2 \ll Q^2$. In this work we define the resolved
contribution by the expression (\ref{neweq}), and in Section 4, we shall
calculate its numerical importance. \par

When the lowest order expression (\ref{neweq}) is resummed, it must
first be removed from the HO corrections. We shall call $HO_s$ the
remaining corrections, and we shall say that the subtraction has been
performed at the scale $M_{\gamma}^2 = Q^2 + E_{\bot 4}^2$.

\section{Results} \hspace*{\parindent} In this section we present the
results obtained with the NLO code described in sections 2 and 3. We
shall compare our predictions with a selected set of H1 data \cite{2r}
and we shall concentrate on a detailed discussion of the various
contributions to the cross section, namely the HO corrections, the
virtual photon structure function contribution and the BFKL-like
contribution. Here we do not intend to perform a complete
phenomenological study of the H1 data \cite{2r,2bisr} that we shall
present in a forthcoming publication. We use the MRST99 (higher
gluon) \cite{a} distribution functions corresponding to
$\Lambda_{\overline{MS}} = 300$~MeV and the KKP fragmentation functions of 
quarks and gluons in $\pi^0$ \cite{b}. The renormalisation and factorization 
scales are taken equal to $\sqrt{Q^2 + E_{\bot 4}^2}$. \par

First we study the cross section $d\sigma/dx_{Bj}$ measured by
H1 \cite{2r} in the range 4.5 GeV$^2 \leq Q^2 \leq$ 15 GeV$^2$ with a
lower bound on the transverse energy, in the $\gamma^*-p$ frame, of the
forward $\pi^0$ given by $E_{\bot 4} > 2.5$~GeV. The HERA proton and
electron beams have laboratory energies 820 GeV and 27.5 GeV
respectively, and the inelasticity defined in (\ref{1e}) is
restricted to the range
$.1 < y < .6$. The forward domain in which the meson is observed is
given by $5^{\circ} \leq \theta_{\pi}^{Lab} \leq 25^{\circ}$ and $x_{\pi}
= E_{\pi}^{Lab}/E_{proton}^{Lab} \geq .01$. Then we
shall consider other kinematical ranges for $Q^2$. \par

\begin{figure}
\centering
\includegraphics[width=4in, height=4in]{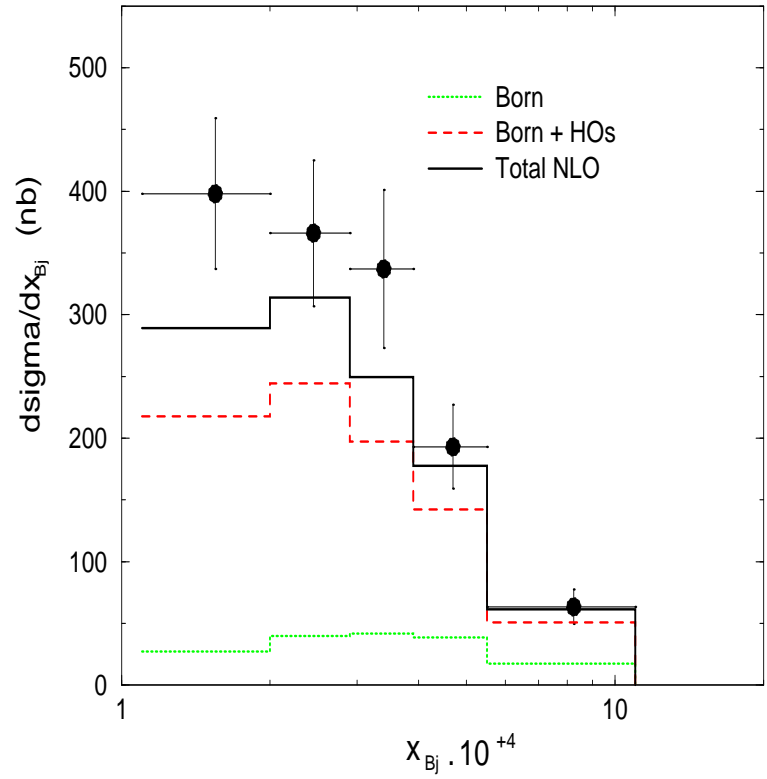}
\caption{Comparison with H1 data in the range $4.5$~GeV$^2 \leq Q^2
\leq 15$~GeV$^2$.}
\end{figure}

The H1 data are
compared to our predictions in Fig.~2. The HO contributions from 
which the resolved
contribution has been subtracted (see the discussion in section 3) are
indicated by $HO_s$. In Fig. 2 the importance
of the HO corrections, especially at small values of $x_{Bj}$ is
dramatically visible. As
discussed below, it is associated with the opening of new channels. \par

The resolved component is non-negligible in this range of $E_{\bot
4}^2$ and $Q^2$. Of course its amplitude depends on the factorization
scale, which we discussed in Section 3. We recall that
the factorization scale used here is $M_{\gamma}^2 = Q^2+ E_{\bot 4}^2$.
It is interesting to look at $<E_{\bot 4}^2>$ and compare it with the value of
$<Q^2>$. According to our calculation $<E_{\bot 4}^2> \simeq
15.3$~GeV$^2$, which does not depend on the value of $x_{Bj}$; $<Q^2> =
6.15$~GeV$^2$ and 10.36~GeV$^2$ for the ranges $1.10^{-4} <
x_{Bj} < 2.10^{-4}$ and $5.5.10^{-4} < x_{Bj} < 11.0 \ 10^{-4}$ respectively. 
These values lead to a virtual photon structure function proportional to
$\log\left ( {Q^2 + E_{\bot 4}^2 \over Q^2} \right )~\simeq~1.25$,
and $0.91$. The dependence of $<Q^2>$ on $x_{Bj}$ also explains
the relative decrease of the resolved component at large $x_{Bj}$.
For two other $Q^2$ ranges we compare H1 data with theory in Fig.~3. We
clearly notice the decrease of the resolved component when $Q^2$
increases and becomes larger than $E_{\bot 4}^2$. \par

\begin{figure}
\centering
\includegraphics[width=2.8in,height=3in]{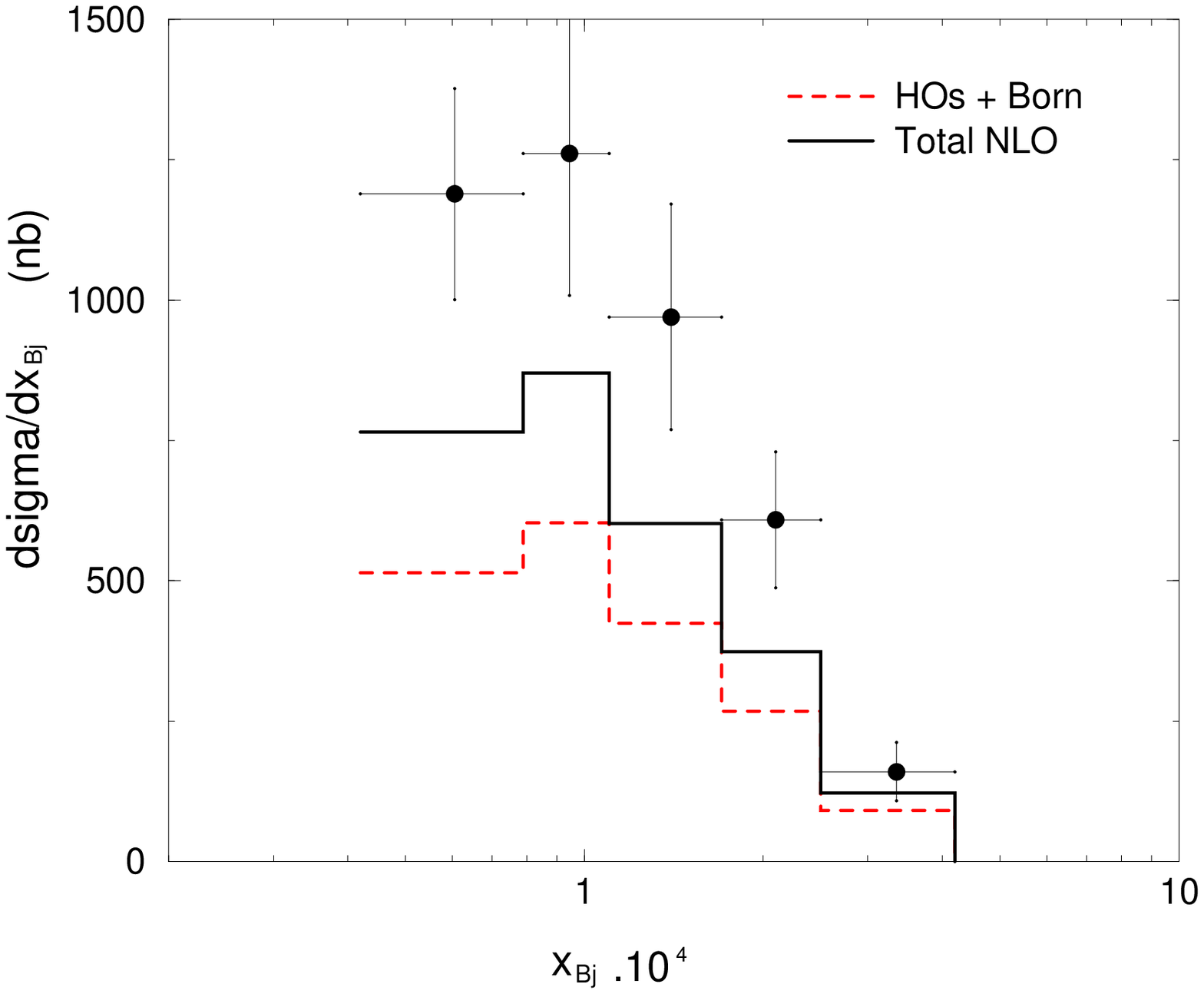}
\includegraphics[width=2.8in,height=3in]{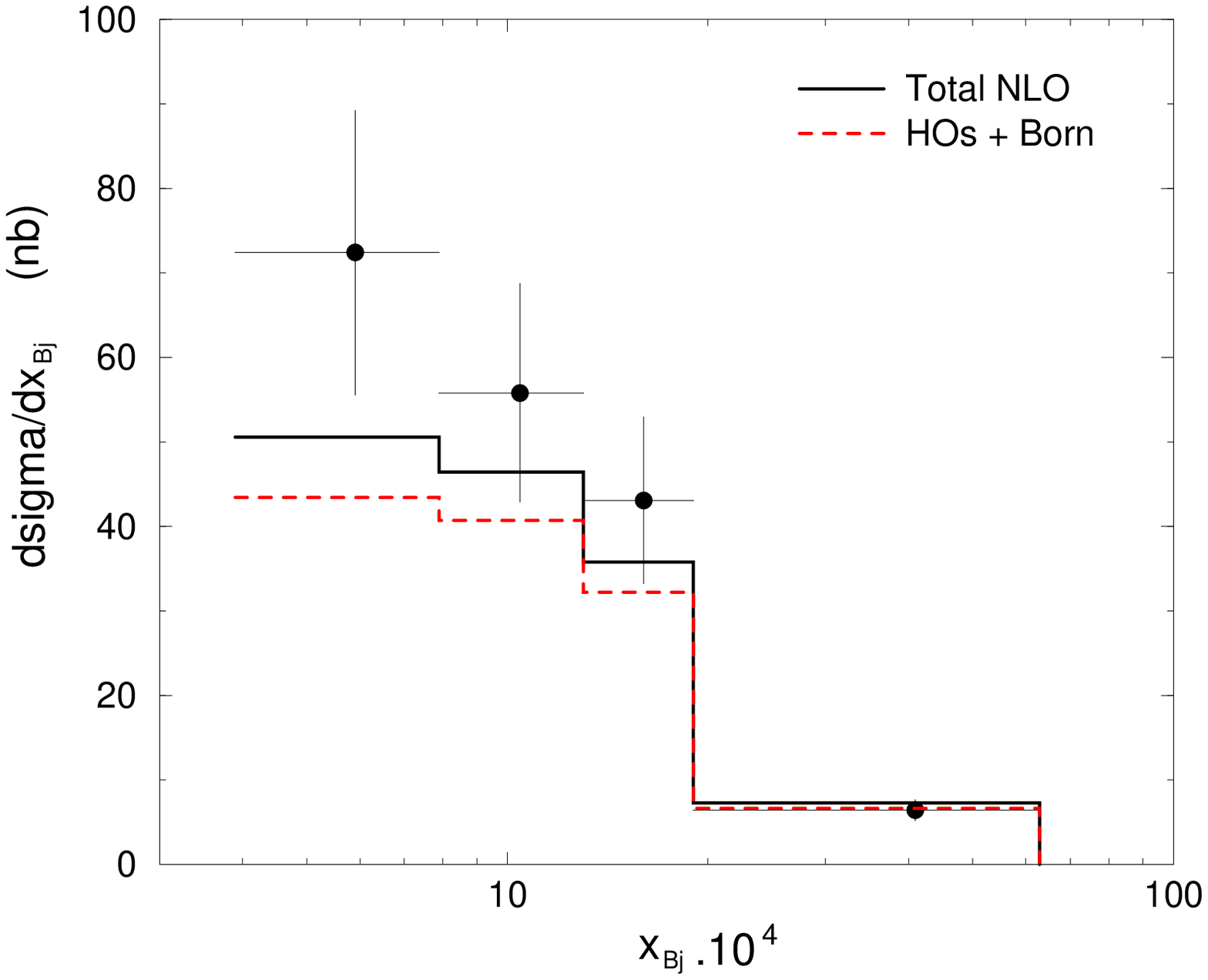}
\caption{Comparisons with H1 data for two $Q^2$-ranges~: $2 < Q^2 <
4.5$~GeV$^2$ (left) and $15 < Q^2 < 70$~GeV$^2$ (right).}
\end{figure}


Let us end this general discussion of our results by noting that 
theory underestimates
data by a small amount. Nevertheless we must keep in mind
that two points are still missing for a more complete comparison. First
we have not yet studied the scale dependence of our results which have
been obtained for the choice $\mu = M = M_F = M_{\gamma} = (Q^2 +
E_{\bot 4}^2)^{{1\over 2}}$, and second, we have not considered HO
corrections to the resolved contribution which are known to be large.
Indeed we can estimate these corrections by using the
Weizs\"acker-Williams approximation (\ref{8e}) which is implemented in
the photoproduction EPHOX code\cite{17r}. From this code we obtain a 
ratio $HO/Born
\simeq 1$ in the kinematical domain corresponding to Fig.~2.\par

\begin{figure}
\centering
\includegraphics[width=4in, height=2in] {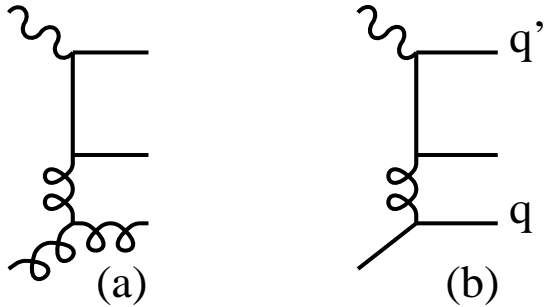}
\caption{Examples of HO diagrams which contribute to the ``BFKL 
Born'' term when the detected hadron is a fragment of the gluon or of 
the quark $q$.}
\end{figure}

Now we turn to a detailed study of the HO contributions. As expected
the two contributions corresponding to the Feynman graphs  
shown in Fig.~4 are
the largest in the forward direction, because of the exchange of a
gluon in the $t$-channel. They also correspond to new subprocesses that
are not
present at the Born level, as soon as the observed partons are the
final gluon or quark $q$~; therefore they do not possess singular collinear
configurations of partons which contribute to the dressing of the
distribution and fragmentation functions already present in the Born
terms. These graphs, with a trigger on the gluon or quark $q$,
also correspond to the Born terms of the BFKL ladder in which extra
gluons are emitted by the $t$-channel gluon. This is precisely the
contribution to the forward cross section \cite{4r,5r} that HERA
experiments H1 and ZEUS should reveal. \par

\begin{figure}
\centering
\includegraphics[width=4in, height=4in]{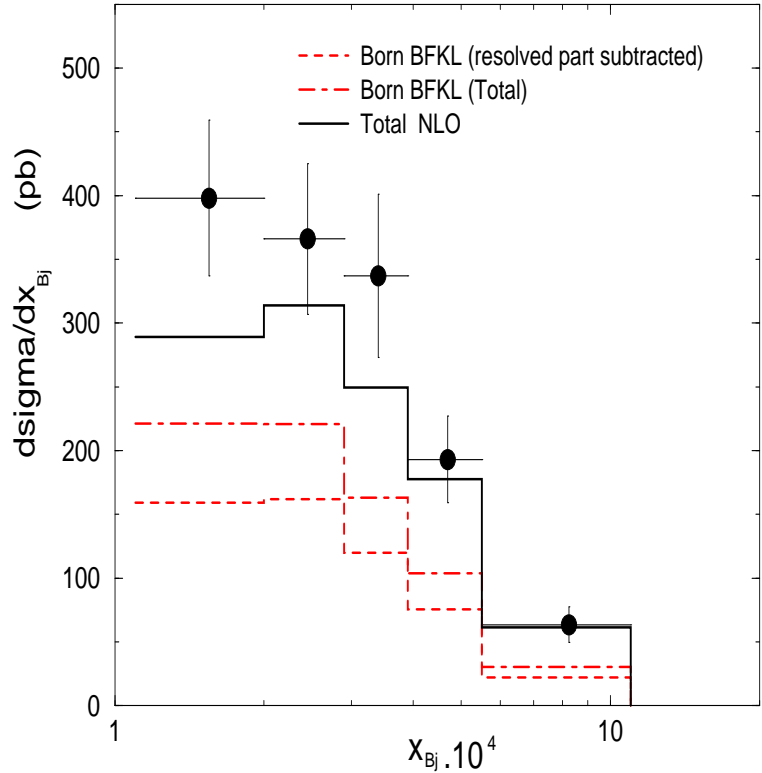}
\caption{Comparison of BFKL Born contributions with the total NLO 
cross section and H1 data for 4.5 GeV$^2 \leq Q^2 \leq 15.0$~GeV$^2$.}
\end{figure}

Figure 5 compares our BFKL Born term contributions and the associated 
resolved contribution with the total NLO
cross section $d\sigma/dx_{Bj}$. These contributions represent more
than two thirds of the total NLO corrections in the small
$x_{Bj}$ region. Actually the BFKL Born result of Fig.~5 also contains
contributions of graphs in which, for instance, the outgoing gluon is
attached to the quark line. However these contributions are expected to
be small as they do not possess the $t$-singularities (in a physical 
gauge) associated with
the exchange of a gluon. To check this point, we calculated the
contributions of the graph shown in Fig.~4b. For a forward trigger on
the quark $q$, we obtain a cross section seven times larger than the
one corresponding to a trigger on the quark $q'$ (for $x_{Bj} \sim
2.10^{-4}$). Therefore we estimate that the part of the curve of Fig.~5
corresponding to the BFKL Born term of Fig.~4 is dominant.\par

\begin{figure}
\centering
\includegraphics[width=4in, height=4in]{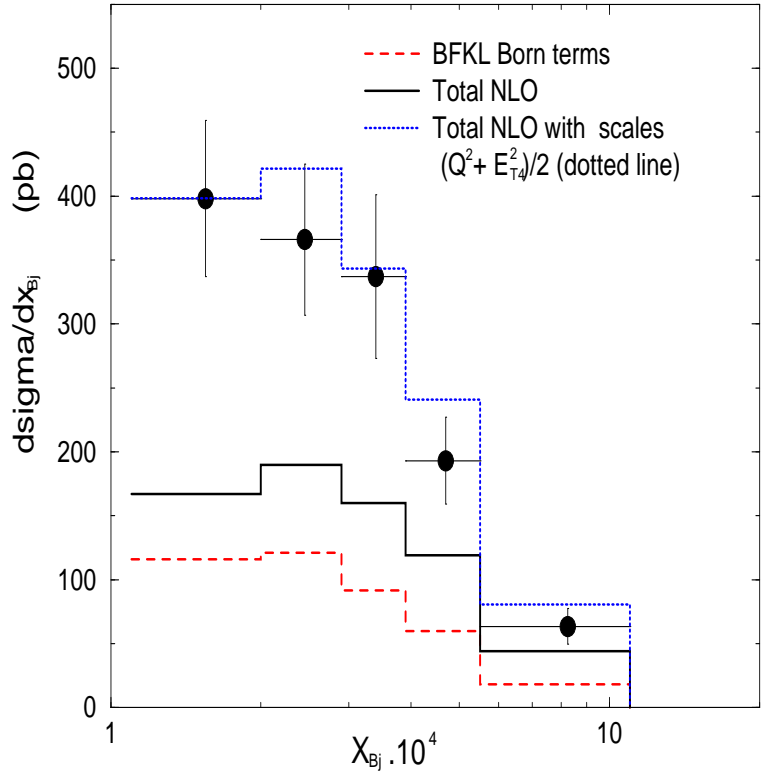}
\caption{The BFKL Born term contribution and the total NLO cross
section with the scales $\mu^2 = M^2 = M_F^2 = p^2_{\bot 4}$. Also shown the
total NLO cross section with scales $(Q^2 + E_{\bot 4}^2)/2$ (dotted
line) and the H1 data ($4.5 \leq Q^2 \leq 15.0$~GeV$^2$).} \end{figure}

  From these results we conclude that the main part of the forward cross
section is due to the BFKL Born terms. Although obtained in the 
course of a NLO calculation, these
terms of order ${\cal O}(\alpha_s^2)$ represent the Born terms of new
channels, namely the $\gamma^*q \to q'\bar{q}^{\, '}q$ and $\gamma^* g \to q
\bar{q}g$ channels. As for any Born terms, we do expect the
contributions of those channels to be strongly dependent on the
renormalization and factorization scales. Therefore, contrary to our
expectations, we are not able to obtain, through our NLO calculations,
a total cross section displaying a weak dependence on the
renormalization and factorization scales. \par

Let us be more explicit by studying the effect of the BFKL
resummation of the small $\log{1 \over x_{Bj}}$ terms which appears
when extra gluons are emitted by the $t$-channel gluon of
Fig.~4 (the BFKL ladder). These contributions have been estimated in
ref. \cite{5r,6r}; they lead to an enhancement of the BFKL Born cross
section by a factor 5 to 10, in obvious contradiction with the
data. However the calculations of ref. \cite{5r,6r} depend on various cuts
and do not include the effect of HO corrections to the leading BFKL
results. These corrections are known to be large; this makes the
leading results not reliable. A more recent approach \cite{8r} includes
a part of these HO corrections in its predictions and finds an agreement
with H1 data \cite{2r}. This last result, obtained with the scales $\mu = M =
M_F = p_{\bot 4}$ ($p_4$ is the momentum of the parton which
fragments into the $\pi^0$), allows us to make a connection with their
approach. Using the same scales we obtain the result of
Fig.~6 for the BFKL Born term and for the total cross section. It is
obvious that there is room for a BFKL-ladder contribution \cite{8r}
between the data and the present theoretical prediction. \par

However the scale $p_{\bot 4}$ is quite large ($<p_{\bot 4}> \simeq
11$~GeV in the H1 kinematics) compared to what is usually used in
large-$E_{\bot}$ reactions. For instance, in the case of $\pi^0$
hadroproduction in fixed target experiments, a scale $M \sim E_{\bot
4}/2$ ($<E_{\bot 4}> \simeq 3.6$~GeV in the H1 kinematics) is used in
ref. \cite{19r} to get an agreement between data and theory. If a
similar scale were used here, we would obtain better agreement between 
data and NLO calculations without any other contributions, as is
demonstrated in Fig.~6 for the choice $\mu^2 = M^2 = M_F^2 = {1
\over 2} (Q^2 + E_{\bot 4}^2)$. It is clear from this discussion that we
cannot accurately determine the importance of a BFKL-ladder component
in H1 data without calculating NLO correction to the BFKL Born terms
considered here or, in other words, without calculating NNLO
correction to the electroproduction cross section.\par

Conclusions similar to the ones of this section have been obtained by 
Kramer and P\"otter
in their study of the forward leptoproduction of jets \cite{(*)}.

\section{Cross section at central rapidity}
\hspace*{\parindent} In this section we study the leptoproduction of
large-$E_{\bot}$ $\pi^0$ in the central region in rapidity (in the
laboratory frame) with the aim to reduce the BFKL Born term
contributions and, consequently, to have a better control of the HO
corrections. Therefore the experimental results obtained in this
kinematical domain should provide good tests of QCD and the possibility
of measuring the virtual photon structure function. \par

Let us consider the following kinematical range which has been explored
by the H1 collaboration in its measurement of the photoproduction of
large-$E_{\bot}$ hadrons \cite{26r}, namely $\sqrt{S_{ep}} = 300$~GeV, $.3
\leq y \leq .7$ and $-1 \leq \eta_{hadron} \leq 1$. We study two lower
limits for the hadron transverse energy, 3~GeV~$\leq E_{\bot}$
and 7~GeV~$\leq E_{\bot}$, in order, as in ref. \cite{27r}, to
estimate the importance of these cuts on the control of the HO
corrections. For $Q^2$, we choose 5~GeV$^2 \leq Q^2 \leq 10$~GeV$^2$
which belongs to the range studied by H1 in the measurement of the dijet
cross section at low $Q^2$ \cite{28r}.\par

\begin{figure}
\centering
\includegraphics[width=2.8in,height=3in]{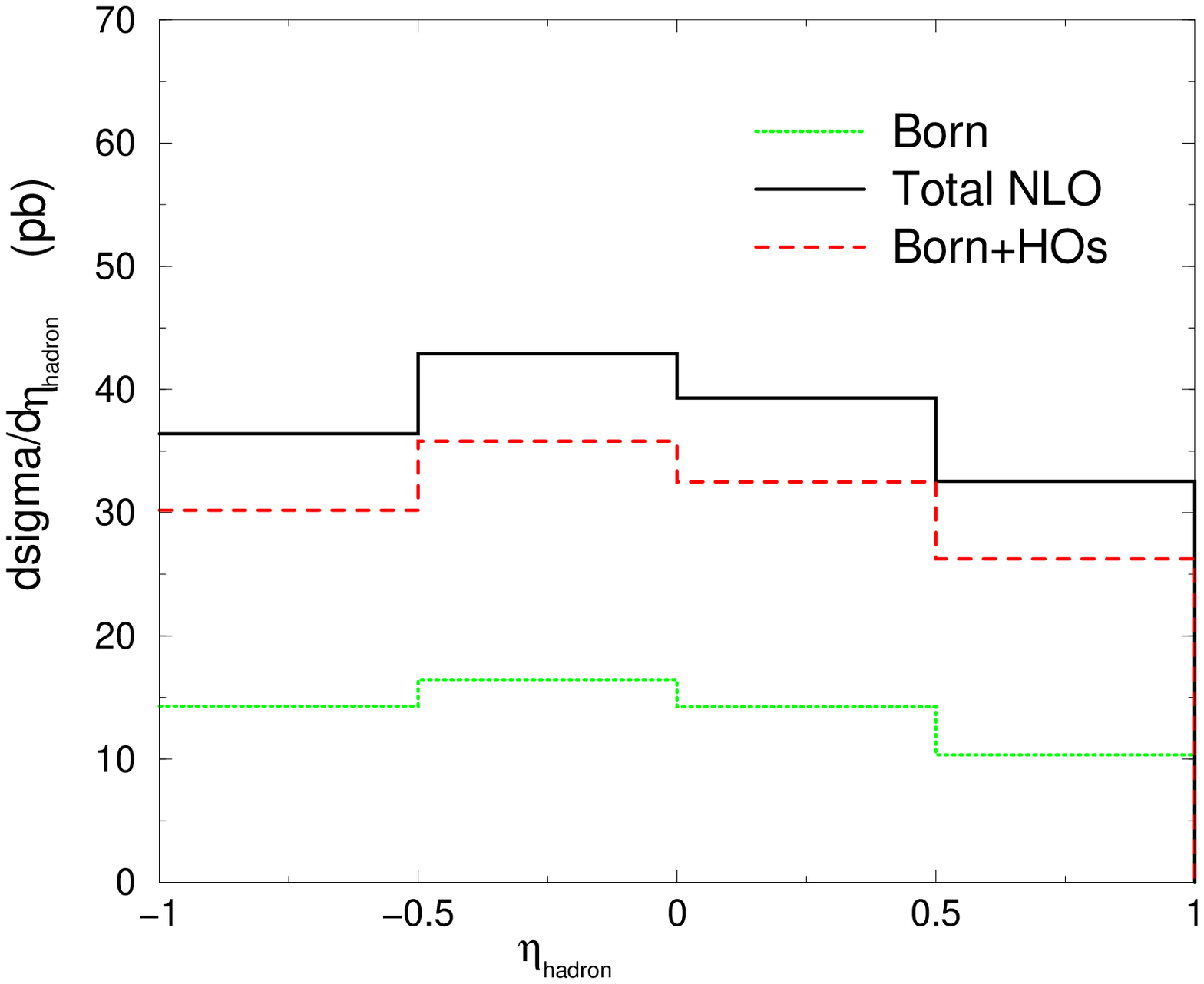}
\includegraphics[width=2.8in,height=3in]{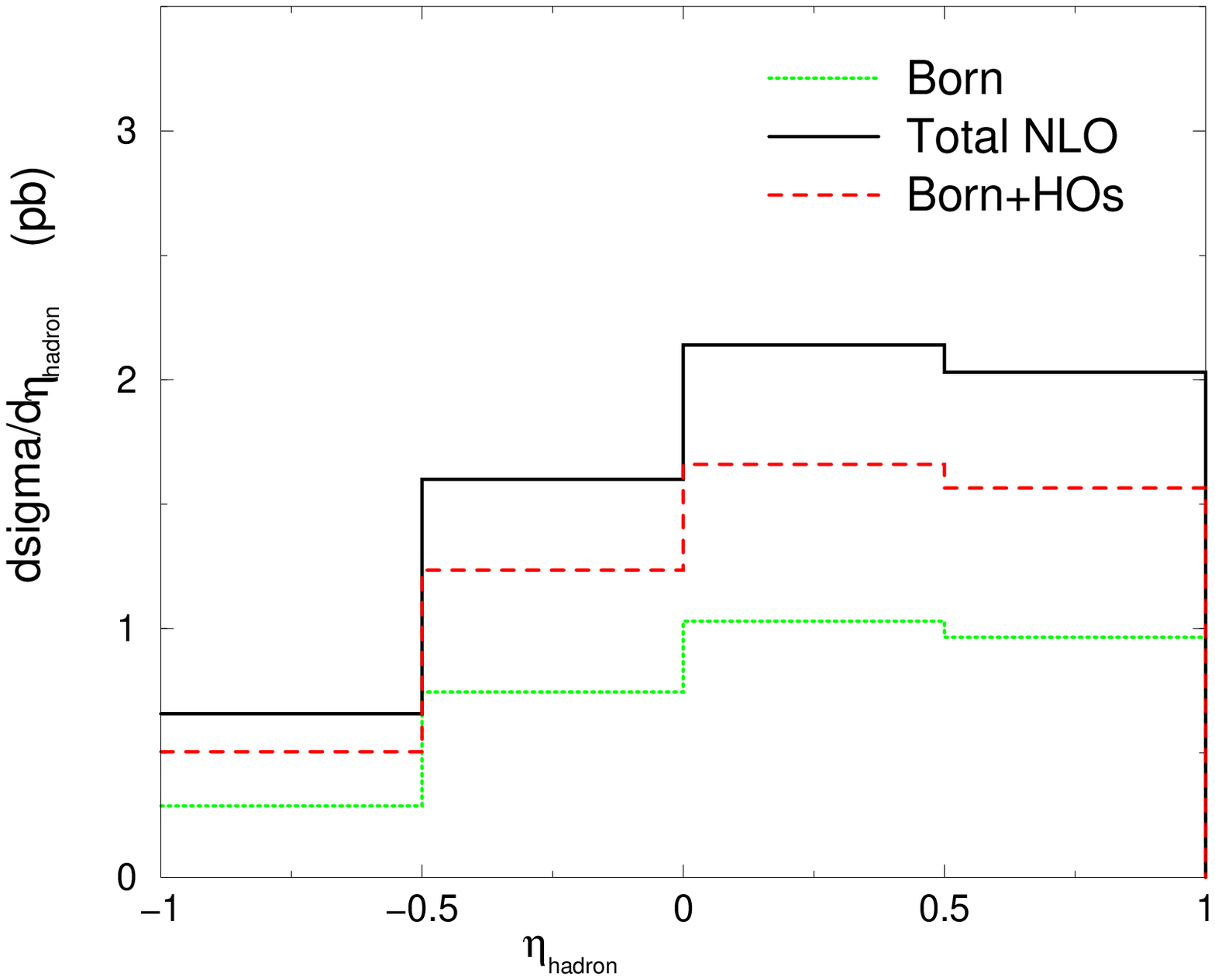}
\caption{The large-$E_{\bot}$ $\pi^0$ cross section integrated over 
$E_{\bot}$ with the cuts $E_{\bot} > 3$~GeV/c (left) and $E_{\bot} > 
7$~GeV/c (right).}
\end{figure}

In Fig.~7 we display the cross section $d\sigma/d\eta_{hadron}$
integrated over $E_{\bot}$ with $E_{\bot} > 3$~GeV (left)
and $E_{\bot} > 7$~GeV (right). All the scales are set equal to
$\sqrt{Q^2 + E_{\bot}^2}$. We clearly see the decrease of the ratio 
$r = {HO_s \over Born}$ when
the cut on $E_{\bot}$ increases, with a value $r \simeq .65$
obtained for $E_{\bot} > 7$~GeV. On the other hand the ratio of
the resolved contribution to the Born contributions increases because
of the larger value of $(Q^2 + E_{\bot}^2)/Q^2$ and reaches a value close to
0.5. One must notice that this ratio is much smaller than in
photoproduction \cite{27r}, since the virtuality of the photon suppresses the
resolved contribution. \par
 
\begin{figure}
\centering
\includegraphics[width=4in, height=4in]{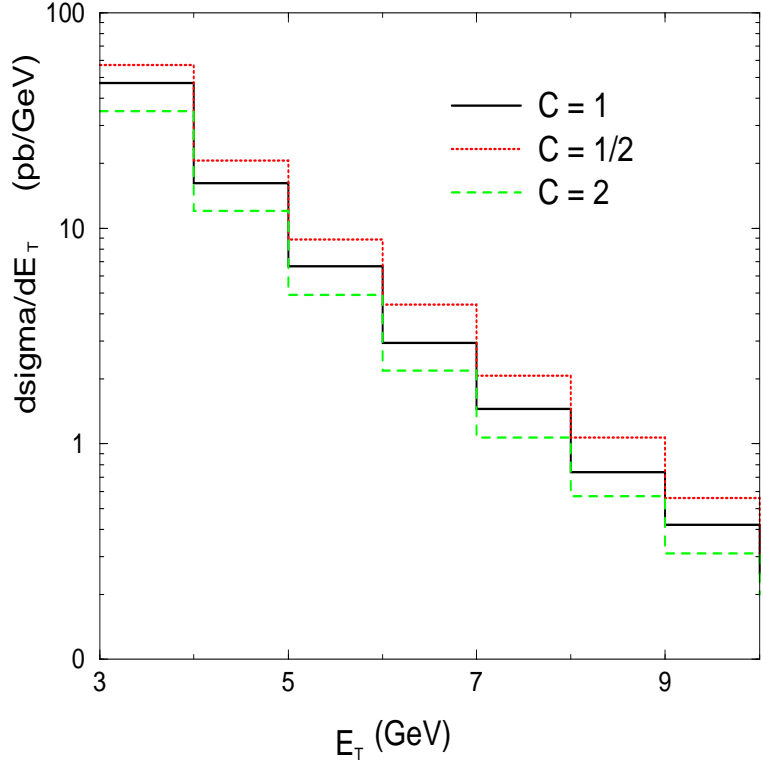}
\caption{The cross section $d\sigma/dE_{\bot}$ for three different 
choices of scales $C\sqrt{Q^2 + E_{\bot}^2}$.}
\end{figure}

We study the scale variation of the cross-section $d\sigma/dE_{\bot}$
in Fig.~8. Three predictions, obtained with the scales $\mu = M = M_F =
C \sqrt{Q^2+E_{\bot}^2}$ and $C = {1 \over 2}, 1, 2$, are displayed. A
change of the scales by a factor 4 results in a change of the cross
section by a factor 2. Therefore, even at large value of $E_{\bot}$, the
cross section is very sensitive to the scale variation.
A similar behavior has been observed in the
photoproduction cross section of large-$E_{\bot}$ hadrons \cite{27r}.

\section{Conclusion}
\hspace*{\parindent} In this work we have carried out the calculation of HO
corrections to the leptoproduction cross section of large-$E_{\bot}$
hadrons. These corrections are implemented in a parton event-generator
offering greater flexibility for the estimation of various observables.
Concerning the numerical importance of these corrections, we have focussed on
two different kinematical domains, namely the central region in
rapidity (in the HERA laboratory frame), and the forward region where
we compare our results with H1 data. \par

In the central region we have found important HO corrections at low
$E_{\bot}$. However these corrections decrease as $E_{\bot}$ increases
with a $K$-factor (HO/(HO + Born)) of about .5 obtained for $E_{\bot} >
7$~GeV/c. In the same range the HO corrections contain a non-negligible
resolved contribution which should allow experiment to constrain the
virtual photon structure function. \par

In the forward region, we find very large HO corrections due to the
opening of new channels related to the BFKL Born terms~; two thirds of
the NLO cross section is due to these contributions. These Born terms,
and consequently the total NLO cross section, are quite sensitive to
scale variations and this forbids any absolute normalization of the cross
section. However one must keep in mind that a good agreement between
the H1 data and the NLO cross section is obtained with renormalization
and factorization scales taken equal to $\sqrt{(Q^2 + E_{\bot}^2)/2}$.
\par

Because of the scale sensitivity, no firm conclusion can be drawn on
the importance of the BFKL resummation in the forward cross section as
measured by the H1 collaboration. Clearly such a study requires the
calculation of NNLO corrections to the leptoproduction cross section.
\\[1cm]
{\bf \large Acknowledgements} We would like to thank J.-Ph. Guillet for 
many useful discussions and suggestions. PA and RB would like to 
thank S. Catani for clarifying many issues regarding the code DISENT
in the preliminary stages of this work. RB and RMG would like to thank 
LAPTH, Annecy and PA and MF would like to thank the Institute of 
Mathematical Sciences, Chennai and the Indian Institute of Science,
Bangalore for their kind hospitality. This work was partially supported by 
CEFIPRA project 1701-1.

\end{document}